\documentclass[conference]{IEEEtran}
\IEEEoverridecommandlockouts
\usepackage{cite}
\usepackage{amsmath,amssymb,amsfonts}
\usepackage{algorithmic}
\usepackage{graphicx}
\usepackage{textcomp}
\usepackage{xcolor}
\def\BibTeX{{\rm B\kern-.05em{\sc i\kern-.025em b}\kern-.08em
    T\kern-.1667em\lower.7ex\hbox{E}\kern-.125emX}}
\begin{document}

\title{A Robust Modular Quantum Processor
}

\author{\IEEEauthorblockN{Ramesh Bhandari}
\IEEEauthorblockA{\textit{Laboratory for Physical Sciences,} 
College Park, Maryland, USA\\
rbhandari617@gmail.com}
}

\maketitle

\begin{abstract}
We explore  the concept of redundancy of critical elements in a quantum computing architecture to circumvent disruption of quantum operations due to a failure of such an element, for example, from a catastrophic cosmic ray event.  We illustrate this concept with reference to a  recently proposed  superconducting modular quantum architecture with a star-like configuration, which has  a  router at the center that enables superconducting qubit interactions across various modules. Regarding this router as a vital element, we propose a double-star configuration, where a loss of one router is backed by the second one. We also examine the usefulness of this double-star configuration under normal conditions, namely, when the quantum hardware has been rendered safe against cosmic rays due to other mitigating actions like shielding or movement to an underground facility. Simultaneous two qubit-pair interactions like two simultaneous CZ gates and multiqubit gates like CCZS are then easily facilitated. 
\end{abstract}


\section{Introduction}
The success of quantum computation depends upon the ability to control and deal with the undesirable effects of external factors such as environmental heating, stray electromagnetic fields, radiation, etc on the state of qubits, which can change,  resulting in what is termed as qubit errors [1].  To mitigate the deleterious effects of such errors on quantum computation, quantum error correcting codes [1,2] like the surface codes [3] have been developed for implementation in quantum computing architectures with the goal to make quantum computation fault tolerant. 

A new, emerging source of concern is  the ubiquitous presence of cosmic rays - muons, gamma rays, protons, neutrons, etc. impinging on earth with high energies - that can have a damaging impact on a quantum device composed of superconducting qubits fabricated on silicon (or sapphire)  [4-9]. One way to reduce the detrimental effects is  through shielding and/or by moving the quantum hardware underground to some depth and even changing its orientation [5,8,9]. Additionally, dynamical techniques to obviate the impact of cosmic rays have also been examined [10], whereby a quantum processor may experience a failure in one of its components due to cosmic rays, but still recover via the application of  error-correcting codes. However, the time lag before the quantum system can fully recover may undermine a running quantum algorithm with a potential loss in accuracy or even its complete failure.

In this paper we invoke the radically different approach of building in redundancy within a quantum hardware, so that if one critical element failed, another acting as a backup would take over to keep continuity of quantum operations,  an idea commonly employed in survivable classical networks [13,14]. In particular, to illustrate the concept of redundancy, we consider the recently engineered star-like architecture [11] in connection with superconducting modular quantum computing.  Fig. 1 illustrates this architecture  (a similar graphical representation derives from [12]).  The central node (router $R_1$) is a capacitor with  four attached switches, each switch connecting the capacitor to the qubit at the other end of the link (see Fig. 2); each switch is made up of a pair of  Josephson junctions [11], which are known to be susceptible to damage by cosmic rays.  The loss of even a single switch (equivalently, a link) would disconnect the associated qubit from interactions with the remaining qubits,   adversely affecting quantum operations; for example,  any ongoing two-qubit interactions through that switch or link in Fig. 1 would cease to exist.
%
To ensure continuous uninterrupted full quantum operations, it then becomes imperative to provide an alternate back-up path to enable interactions across different pairs of qubits. In this context we propose  a new architecture depicted in Fig. 3, which we call the double-star configuration.

In Section II we list the main features and the mechanism of the star-configuration [11]. Section III discusses the extension to the double-star configuration containing an extra (backup) router. This new configuration provides other advantages under normal conditions when the quantum hardware is assumed to have been rendered completely safe against cosmic rays through other means such as shielding and/or placement underground.  In this situation, the second router is active (not dormant) and together with the first router enables simultaneous two-qubit operations such as the CZ gates as well as multiqubit  operations such as the  CCZS gates. Section IV is the summary with discussion. 

\section{Current Configuration}
Fig.1 is a graphical representation of the four superconducting qubits arranged to interact with each other [11]; the details of this arrangement are below:
\begin{figure}[htbp]
\centerline{\includegraphics{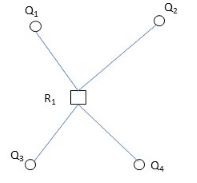}}
\caption{The current configuration where a number of qubits (shown four here) interact with each other through the router $R_1$, one qubit pair at a time; router $R_1$ is composed of a capacitor and four switches, each switch connecting the capacitor to the qubit at the other end of the link (see Fig. 2); a switch, however, can be  turned off to disconnect the associated qubit from the rest of the qubit network, thus preventing any two-qubit interaction between this qubit and the remaining three qubits (see text for more details).}
\label{fig}
\end{figure}
\begin{figure}[htbp]
\centerline{\includegraphics{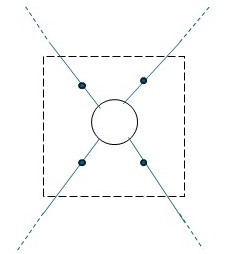}}
\caption{The dashed square shows an expanded view of the router $R_1$ of Fig 1; the solid dots are the ON-OFF switches, each connected to a central capacitor (solid circle) at one end and to a qubit of Fig. 1 at the other end; the switches are made up of Josephson junctions in SQUID formations [11].}
\label{fig}
\end{figure}
\begin{itemize}
\item Superconducting qubits, $Q_i, (i=1.2,3,4)$, can be connected to each other via the  router $R_1$ in a star-like configuration; theoretically, any number of qubits (not just four) may be connected through this router; interactions occur pair-wise, one pair at a time.
\item Between the qubit and the router on each link, there is an ON-OFF switch for the associated qubit (say $Q_1$) to control interaction with one of the remaining qubits ($Q_2, Q_3, Q_4$); the architecture of the router is shown schematically in Fig. 2. The switch comprises a superconducting quantum interference device (SQUID), composed of two Josephson junctions in parallel, whose induced inductance L is varied by an external flux threading the SQUID; the OFF state arises when L is large and the ON state when L is small. 
\item When all qubit links  are in the OFF state, all possible qubit pair interactions are excluded and only single qubit gates are possible.
\item Two qubits interactions are enabled when two switches are in the OFF state; a total of six qubit pair interactions are possible ($Q_1Q_2, Q_1Q_3,Q_1Q_4, Q_2Q_3, Q_2Q_4, Q_3Q_4$), one pair at a time.
\end{itemize}

\section{New Configuration}
\begin{figure}[htbp]
\centerline{\includegraphics{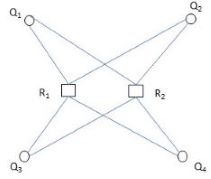}}
\caption{The double-star configuration that has two routers, $R_1$ and $R_2$, to which all the four qubits have direct links (ON-OFF switches exist on each link as before; see Fig. 2); one router (say $R_1$) is an active router while the other router ($R_2$) is inactive and serves as a backup; if $R_1$ fails, then $R_2$ takes over; under normal conditions,  router $R_2$ is also active, enabling simultaneous two-qubit operations and three-qubit gates (see text for more details).}
\label{fig}
\end{figure}

\subsection{Survivable Configuration}
Fig. 3 shows the new proposed architecture,  where we have an additional router $R_2$ to which all the four qubits of the modular architecture of [11] are assumed connected on a different set of links through ON-OFF switches, one on each newly-added link, as before. This configuration, a double-star configuration, can act as a survivable configuration in the following way: the part of the network centered on $R_2$ is initially set to a dormant state by placing the switches on the four-incident links, $Q_1R_2, Q_2R_2, Q_3R_2$, and $Q_4R_2$ to the OFF position. All the two-qubit interactions, one pair at a time, occur through the router $R_1$ as in Fig. 1. If the router $R_1$ fails, a local fast signaling mechanism engineered into the network (signaling time much less than a typical gate operation time) then activates  immediately the part of the network through the router $R_2$,  thus preserving the continuity of an ongoing quantum computation; for example, an intended two qubit interaction between qubits $Q_1$ and $Q_3$  can now take place through the functional  router $R_2$ (note that the failure of the router $R_1$ here means  the loss of one or more  links incident on  it).

\subsection{Simultaneous Two-qubit and Multiqubit Interactions}
Here we consider the case where  the damaging impact of cosmic rays on the quantum device has been eliminated  through appropriate shielding and/or by locating the quantum device underground at a  suitable depth. Under this assumption,, we can  keep both the routers in Fig. 3 active. This then permits  simultaneous quantum operations pair-wise among the qubits accessible through the two routers without any interference effects; for example, two CZ gates,  one between qubit $Q_1$ and qubit $Q_4$ and the other between qubit $Q_2$ and $Q_3$  can occur simultaneously through the two available and physically separated routers; the $Q_1Q_4$ interaction can proceed through $R_1$ (or $R_2$) and the $Q_2Q_3$ interaction through $R_2$ (or $R_1$).   Architectures that allow for simultaneous operations are always desirable as they speed up quantum operations. In [11], the single CZ gate was demonstrated with a fidelity of 96\%; consequently,  each of the two  simultaneous CZ operations in Fig. 3 would very likely have  the same level of fidelity (in [11], preliminary results for two simultaneous two-qubit operations through the same router $R_1$ are cited for the SWAP gate with a somewhat reduced level of fidelity in comparison to the single CZ gate; they achieve this by detuning two-qubit operation frequencies for different pairs of qubits, but we feel that this process may become more cumbersome due to the number of qubit pairs increasing with the number of qubits incident on the router).

Examining Fig. 3 further, qubit $Q_1$ could also be coupled to qubit $Q_3$ through $R_1$ (or $R_2$) and qubit $Q_2$ to qubit $Q_4$ through $R_2$ (or $R_1$) simultaneously. In fact, there is a total of 15 possible double-pairs to choose from (this follows from the simple combinatorics of selecting  two different pairs from a total of six given pairs); of the 15 possible double-pairs, three are of the form: ($Q_iQ_j, Q_kQ_l)$, where $i,j,k,l$ are all different, e.g., ($Q_1Q_3, Q_2Q_4$), allowing for two simultaneous CZ gates as an example;  
the remaining 12 have one index from each pair equal, e.g., ($Q_2Q_3, Q_2Q_4$); such configurations  allow for three qubit (e.g., $Q_2Q_3Q_4$) gate formations like the CCZS (controlled–CZS) gate through the simultaneous application of two CZ gates, both involving the common control qubit $Q_2$ in the above three-qubit example; the CCZS gate applies both the CZ and the SWAP gates to the target qubits $Q_3$ and $Q_4$ conditioned on the control qubit $Q_2$ [15].

While all quantum algorithms can be expressed in terms of the universal set of single and two-qubit gates [1], the existence of multiqubit gates like the above discussed  three-qubit CCZS gates in Fig. 3 circumvents the need for such decomposition into the universal set when quantum algorithms require  multiqubit entanglement gates; rendering  multiqubit gates into the basic universal set only increases the hardware overhead and should be avoided; for example, to implement the three qubit Fredkin gate, five two-qubit gates are required [16]. 

Extending the double-star configuration to a multi-star configuration by adding additional routers in a similar way (e.g., a single router addition to Fig. 3 creates a triple-star configuration) would likewise enable multiqubit gates beyond the three-qubit CCZS gate discussed here.

\section{summary and discussion}


In summary, the paper is an illustration of how redundancy of critical elements in a quantum hardware can enable continuity of quantum operations in the event of their failures due to  significant high energy cosmic ray events (or in the most general sense, any other cause). The general concept of redundancy and its utility  in quantum architectures are examined  with respect to a modular superconducting quantum processor, a current important area of research  as quantum architectures scale to accommodate  higher number of physical qubits required for  practical quantum computation problems of interest.   For the considered modular quantum processor [11], we focused on the central router as a critical element and propose a  double star configuration formed with a backup router.  In situations, where the probability of failure due to cosmic rays or any other cause is made negligible, this configuration then serves as an easy means to simultaneous two-qubit operations like the CZ gates without loss in fidelity compared to a single CZ gate and multiqubit operations like the CCZS gate.  Such simultaneous operations and multiqubit gates speed up quantum computations and reduce quantum hardware requirements, desired features of a quantum computing platform. 





\color{red}%

\end{document}